\documentstyle[aps,prl,twocolumn,epsf,floats]{revtex}
\newcommand{\beq}{\begin{equation}}
\newcommand{\eeq}{\end{equation}}
\newcommand{\bea}{\begin{eqnarray}}
\newcommand{\eea}{\end{eqnarray}} 
\newcommand{\veps}{\varepsilon} 

\begin{document}




\title{ Dilute Quantum Droplets }

\author{ Aurel BULGAC }

\address{ Department of Physics,  University of Washington,
Seattle, WA 98195--1560, USA  }

\date{\today }

\maketitle

\begin{abstract}

In the limit when the two--body scattering length $a$ is negative and
much larger than the effective two--body interaction radius the
contribution to the ground state energy due to the three--body
correlations is given by the Efimov effect. For particular values of
the diluteness parameter $\rho |a|^3$ the three--body contribution can
become the dominant term of the energy density functional. Under these
conditions both Bose--Einstein and Fermi--Dirac systems could become
self--bound and either boson droplets or fermion ``designer nuclei''
of various sizes and densities could be manufactured.

\end{abstract}

\vspace{0.5cm}

{PACS numbers: 03.75.Fi, 05.30.Jp}


\vspace{0.5cm}

\narrowtext

The extraordinary progress in realizing Bose--Einstein Condensates
(BEC) and of the Fermi--Dirac counter parts in traps \cite{bec,marco}
raises the question of whether they can exist also in isolation,
without a trap potential. In the dilute limit, the physical properties
are entirely determined by the two--particle scattering length $a$. To
leading order in $a$ the energy per particle of a uniform condensate
is given by
\beq
E_2 =  \frac{1}{2}g_2\rho ,  
\quad g_2 = \frac{4\pi \hbar^2 a}{m}
\label{eq:en2}
\eeq
where $\rho$ is the number density and $m$ is the particle mass. For
positive $a$ the energy is positive, suggesting that there is no
binding. The energy density, including higher order corrections, is given by
\bea
{\cal{E}}_2(\rho ) &=& \frac{g_2\rho^2 }{2}
\left \{ 1 +\frac{128}{15\sqrt{\pi}} \sqrt{\rho a^3} \right . \nonumber \\
&  + &
\left . \left [\frac{8(4\pi-3\sqrt{3})}{3}\ln (\rho a^3) + C \right ]
\rho a^3 + \dots \right \} , \label{eq:enden} 
\eea
where $C$ is a constant \cite{bs,dick,braaten}.  These higher order
corrections depend on dimensionless parameter $\rho a^3$ up to the
term $C$, which depends on the range of the two--particle interaction.
A similar expression to Eq. (\ref{eq:enden}), but for the energy of a
dilute Fermi system, in the case of positive scattering lengths, was
derived essentially in parallel with the Bose case
\cite{bs,dick}. The positive energy state is stable with
respect to long wave density fluctuations, but is only metastable if
the underlying two--particle interaction supports bound states. In
fact most of the BEC decay into the two--particle bound states, but
they are still quite long lived.  The decay rate of the condensate
depends on $a$ as $\Gamma \propto a^4 \rho ^2$, see Refs.
\cite{recomb} and earlier references cited therein for details and
some rather unexpected features.  In contrast, when the effective
scattering length is attractive ($a<0$), the energy expression
Eq. (\ref{eq:en2}) is unbounded from below and the uniform density
state is unstable with respect to density fluctuations. The time scale
here is very short ($\Gamma \propto \sqrt{|a|\rho}$). It would thus
seem that a droplet of BE condensed matter would be either unbound or
would decay very quickly.  This is not necessarily the case. The
picture is radically changed when one considers the effects of the
three--body correlations.

Recently it was confirmed experimentally the fact that one can now
control the two--body scattering length by placing the atoms in an
external field \cite{fesh}, see also Refs. \cite{el}. For those cases
where successful Bose--Einstein condensation was achieved, the bare or
unadulterated by any external fields two--body scattering length $a$
was comparable in magnitude with the two--body potential radius. More
exactly, the value of the scattering length is comparable with the van
der Waals length \cite{braaten,gf}.
A new regime is now attainable, when the Bose system is still dilute,
but with respect to a new characteristic length.  If in the two--body
system the scattering length is very large, i.e. if $|a|\gg r_0$,
there are now two independent dimensionless parameters, $\rho a^3 $
and $\rho r_0^3$.  In this limit in the three--body system one
has a very unusual phenomenon, the so called Efimov effect
\cite{efimov}, which is manifested in the appearance of a very large
number of three--body bound states $N\approx s_0/\pi\;\ln (|a|/r_0)$,
where $s_0\approx 1.0064$, all with the same quantum numbers
$0^+$. The spatial extensions of these Efimov states range from
distances of the order of $O(r_0)$, for the tightest bound one, to
distances $O(a)$, for the least bound state. These three--body bound
states appear irrespective of whether the two--body scattering length
is positive or negative. The sizes of these states change from one
state to the next by a factor $\exp ( \pi/s_0)\approx 22.68$, while
their energies change by a factor $\exp ( -2\pi/s_0)$. The spectrum
has an exponential character and the bound state wave functions obey a
simple scaling law. The properties of these states easily follow from
the fact that in the region $r_0 \ll R\ll |a|$, where
$R^2=2(r_{12}^2+r_{23}^2+r_{31}^2)/3$ and $r_{kl}$ is the distance
between particles $k$ and $l$, ($k,l=1,2,3$), there is an effective
three--particle interaction with an universal attractive character
$-s_0^2\hbar^2/2mR^2$. The universality of this effect resides in the
fact that all its properties are fully determined by the scattering
length $a$, the radius of the two--body interaction $r_0$, the
logarithmic derivative $\Lambda$ of the three--body wave function at
$R\approx r_0$ and the dimensionless constant $s_0$.  If the particles
have spin, the situation can become a little bit more complicated, as
the Efimov states can appear for various values of the total spin of
the three--body system. Typically the spectrum ceases to be strictly
exponential and the wave functions have a somewhat more complex
structure.  However, on average the qualitative features of these new
type of Efimov three--body states remain largely unchanged
\cite{aurel}.

Efimov \cite{efimov3} has also shown that at zero energy the amplitude
for the three--particle collisions is determined by the same universal
interaction $-s_0^2\hbar^2/2mR^2$. In the case of negative scattering
lengths $(a<0)$ this amplitude is given by the following universal
formula
\beq
g_3 =\frac{12\pi \hbar^2 a^4}{m} 
\left [ d_1 +d_2\tan \left ( s_0 \ln \frac{|a|}{|a_0|}+
\frac{\pi}{2} \right ) \right ] , \label{eq:M}
\eeq
where the numerical values of the universal constants $d_1$, $d_2<0 $
and $a_0$ have been determined numerically recently by Braaten {\it et
al.}  \cite{eric}.  Here $a_0$ is the value of the two--body
scattering length for which the first three--body bound state with
zero energy is formed. In the present parametrization $a_0$ replaces
the logarithmic derivative and the matching radius used by Efimov
\cite{efimov3,eric}.  Unlike $d_1$ and $d_2$, the parameter $a_0$ is
system dependent and is also a genuine three--body
characteristic. When approaching the three--body threshold (by
changing the strength of the two--body interaction), just before the
three--body bound state is formed, $g_3\rightarrow -\infty$, in
complete analogy with the behavior of a two--body scattering
amplitude. After the three--body state has appeared $g_3$ is positive
and $g_3\rightarrow \infty$ when the threshold for the appearance of
the three--body bound state is approached from the other side.

If the three--body zero--energy scattering amplitude $g_3$ is known,
the contribution of the three--particle collisions to the ground state
energy density of a dilute Bose gas can be evaluated in a similar
manner as the leading order contribution of the two--body collisions
\beq
{\cal{E}}_3(\rho ) =  \frac{1}{6}g_3\rho ^3 .
\eeq
The experience gained from studying corrections to the main two--body
contribution to the ground state energy of a dilute quantum gas, see
Refs. \cite{bs,braaten,dick} and Eq. (\ref{eq:enden}), shows that all
other contributions are small in the dilute limit and they are not
expected to lead to any qualitative changes in the properties of these
systems. (In the case of fermion systems pairing correlations however
could lead to a significant rearrangement of the ground state
properties, but energetically the correction is typically small and
the density profiles are not modified in any drastic manner,
see e.g. nuclei \cite{ring}.) The ${\cal{E}}_3(\rho )$ contribution to
the energy density can dominate the contribution arising from the
two--body collisions if the argument of the tangent is close to
$(2n+1)\pi/2$ and the scattering length $a\approx a_0 \exp
(n\pi/s_0),\quad n=0,1,\dots$ is such that a three--body state is on
the verge of appearing or it has just been formed.  The situation is
somewhat unique in this limit.  With respect to the two--body
collisions the system is extremely dilute, but somewhat less dilute
with respect to three--body collisions.

At the points where a three--body bound state appears, where $g_3$
becomes infinite and in the immediate neighborhood of them, the
present calculational scheme fails, since the contribution of the
three--body collisions has been evaluated only in the leading order of
the gas approximation with respect to the three--body collisions.  In
the region near such poles, the contribution of the three--particle
collisions could dominate over the genuine two--particle
contributions, for appropriate values of $\rho$ and $a$.  In a trap
the density profile of such a trapped Bose--Einstein condensed gas is
given now by the new ``Thomas--Fermi'' formula
\beq
\rho (\bbox{r}) = 
\sqrt{ \frac{2[ \mu -V_{ext}(\bbox{r})]}{g_3} 
 +\left ( \frac{g_2}{g_3}\right ) ^2} -  \frac{g_2}{g_3},
\eeq
where $V_{ext}(\bbox{r})$ is the trapping potential and $\mu $ is the
chemical potential. Besides obvious changes in the density profile of
a trapped Bose--Einstein condensed gas, the spectrum of the elementary
and collective excitations is, naturally, modified as well, as the
compressibility of such a system is significantly affected by the
three--body contribution to the ground state energy ${\cal{E}}_3$.  I
would like to note here that the analyses of Refs. \cite{g3} consider
a formally similar situation, but with a fictitious repulsive
three--body force, whose nature and strength are never specified.

Since the two--body contribution to the ground state energy of a
dilute Bose gas is negative, the three--body collisions in the regime
where $g_3>0$ could lead to the stabilization of the system.  What is
particularly interesting for such a system is that a boson droplet --
a boselet -- could become self--bound and the trapping potential is
not required anymore to keep the particles together.  In the absence
of the trapping potential and for a very large number of particles a
boselet will have an almost constant density, corresponding to the
infinite matter equilibrium density
\beq
\rho _0 = -\frac{3 g_2}{2g_3}.
\eeq
The ground state energy of an ensemble of $N$ bosons in the meanfield 
approximation is given by \cite{scaleout}
\bea
E(N) & = &  \int d^3r \veps (\bbox{r}) = \int d^3r 
    \left [
\frac{\hbar^2}{2m}|\bbox{\nabla}\psi(\bbox{r}))|^2  
    \right . \nonumber \\
& & 
    \left .
+ \frac{1}{2}g_2\rho(\bbox{r})^2 + \frac{1}{6}g_3\rho(\bbox{r})^3
    \right ],
\eea
where $\veps(\bbox{r})$ is the energy density and
$\rho(\bbox{r})=|\psi(\bbox{r})|^2$ is the number density.  The
density profile of semi--infinite matter can be shown to be given by
\beq
\rho (z) = |\psi (z)|^2= \frac{\rho_0}{1+\exp (2\kappa_0 z)},
\eeq
where 
\beq
\kappa _0 =  \sqrt{ \frac{2m|\mu_0|}{\hbar^2} },\quad 
\mu _0    = -\frac{ 3g_2^2}{8g_3}, 
\eeq
$z$ is the spatial coordinate normal to the surface and $\mu_0$ is the
chemical potential corresponding to infinite matter at equilibrium
density $\rho_0$.  One can now determine the surface tension $\sigma $
of boselets from the obvious relation
\beq
\sigma  = \int _{-\infty}^{\infty} dz [\veps(z)-\mu_0\rho(z)]
= \frac{g_3\rho_0^3}{12\kappa_0}.
\eeq
The spectrum of both volume and surface sound waves can then be easily 
specified. The density profile of an infinite slab of finite width has 
a simple expression as well
\beq
\rho (z) = \rho _0\frac{\mu}{\mu_0} \frac{1}{1+
  \displaystyle{\sqrt{1-\frac{\mu}{\mu_0} } \cosh (2\kappa  z )} } ,
\eeq
where $\mu = -\hbar ^2 \kappa^2 /2m $ and the chemical potential
satisfies the restrictions $ \mu_0\le \mu < 0$. It is straightforward
to show that for any slab with a width larger than its skin thickness,
the quantities $(\mu-\mu_0)/\mu_0 $ and $(\kappa_0 - \kappa)/\kappa_0$
are both exponentially small. These facts (along with numerical
evidence not presented here) suggest that the basic properties of a
$N$--particle boselet are given by the following relations
\bea
& & \rho (\bbox{r})  \approx \rho_0 
 \left ( 1 + \displaystyle{ \frac{1}{2\kappa R} }\right ) \frac{1}{ 1+
 \displaystyle{\frac{\cosh (2\kappa r)}{\cosh (2\kappa R)}}  },\\
& & R = r_0 N^{1/3}+ r_1N^{-1/3} + {\cal{O}}(N^{-2/3}), \\
& & E(N)=\mu_0 N + 4\pi r_0^2 \sigma N^{2/3} + {\cal{O}}( N^{1/3}) ,\\
& & \omega^2_l=\frac{\sigma l(l-1)(l+2)}{m\rho_0R^3}+{\cal{O}}( N^{-4/3}),
\eea
where $r_0=(3/4\pi\rho_0)^{1/3}$, $\mu =dE(N)/dN=-\hbar^2\kappa^2/2m$
and $\omega_l$ is the frequency of the surface vibration mode with
angular momentum $l$.  The absence of the constant term in the
expression for the radius was established numerically. The central density 
is larger then $\rho_0$ due to surface tension.

The possibility that the entire system can also undergo a transition
to a gas phase of trimers cannot be ruled out at this time.  Since
there are no Efimov states for four or more particles this trimer
phase is perhaps unique. One cannot fail to see here an analogy with
the Cooper pair--BEC crossover in the fermion case \cite{mohit}.  At
each new three-body threshold, when $g_3$ becomes infinite, a new
trimer phase appears, made of spatially larger trimers.  The density
drops naturally by a factor of 3 if a trimer phase is formed.  The
trimer--trimer scattering length is expected to be of the order of the
trimer size, i.e. of order $a$, but the sign of this trimer--trimer
scattering amplitude is so far unknown.  Upon collapsing into trimers
the interaction energy decreases significantly, as now this energy is
controlled by effective two--body processes only. If a trimer phase is
formed, the size of the cloud in the trap should change abruptly.

It is obvious that in the case of a Fermi--Dirac system the role of
three--body collisions could play an analogous role, if the Efimov
effect could take place. Since three identical fermions could never be
all in a relative $s$--state, the particles should have either a spin
larger than 1/2 or some other additional degree of freedom, like
isospin in the case of nucleons.  The characteristics of the Efimov
effect for the case of particles with arbitrary spin and/or other
discrete degrees of freedom (as well as arbitrary masses) have been
established in Ref. \cite{aurel}. Let me consider for illustrative
purposes the case of two identical spin--1/2 fermion species $\pi$ and
$\nu$.  In the meanfield approximation the ground state energy can be
evaluated using the following energy density functional
\bea
& & E(N_\pi,N_\nu) =
\int d^3r \left \{  
\frac{\hbar^2}{2m}[\tau_\pi(\bbox{r}) + \tau_\nu(\bbox{r}) ]
                               \right .\nonumber \\
& &  
+ \frac{g_2}{2}\left [\rho_\pi(\bbox{r})  +\rho_\nu(\bbox{r})   \right ] ^2
- \frac{g_2}{4}\left [\rho_\pi(\bbox{r})^2+\rho_\nu(\bbox{r})^2 \right ] 
\nonumber \\
& & \left .
+\frac{g_3}{4} \rho_\pi(\bbox{r}) \rho_\nu (\bbox{r})
[ \rho_\pi(\bbox{r}) + \rho _\nu (\bbox{r}) ] \right \} , 
\label{eq:fermilet}
\eea
where $\rho_{\pi,\nu}(\bbox{r})$ are the corresponding number density
distributions, $N_{\pi,\nu}= \int d^3r\rho_{\pi,\nu}(\bbox{r})$ and
$\hbar ^2 \tau _{\pi,\nu}(\bbox{r})=\hbar ^2 \sum _{n=1}^{N_{\pi,\nu}/2}
2 |\bbox{\nabla} \psi _{n}^{(\pi,\nu)}(\bbox{r})|^2$ are the momentum
distributions, defined through the corresponding single--particle wave
functions $\psi _{n}^{(\pi,\nu)}(\bbox{r})$.

Again, the most interesting regime to consider is that of a negative
two--body scattering length $g_2<0$.  If the three--body term is
repulsive, that is if $g_3>0$, a dilute Fermi--Dirac droplet will
behave very much like a nuclear system. In this case the two--body
effects will have an attractive role and the energy density functional
of such a system will have the same qualitative structure as the
popular Skyrme energy density functional in nuclear physics
\cite{skyrme}. The Fermi--Dirac droplets -- the fermilets -- will have
entirely unexpected properties: they will be self--bound and show
saturation properties as well, see Fig.  1. The existence of an
equilibrium state for infinite homogeneous fermionic matter is a
sufficient condition for the existence of finite systems as well, see
the case of boselets for example. Since one would be able to change
the relative magnitude of the two--body and three--body interactions
the central density of a fermilet can be controlled. In the absence
of Coulomb interaction the number of particles in a single fermilet is
arbitrary.  A particularly interesting aspect could be the interplay
between the formation of Cooper pairs and fermionic trimers, since
when $g_3>0$ a trimer bound state exists.  The possibility to have a
Fermionic system, with attractive two--body effective interactions and
repulsive three--body effective interactions, opens thus the way
towards the creation of ``designer nuclei'', an almost unthinkable
flexibility, which could not be matched even by atomic clusters.
Self--bound droplets of mixtures of fermions and bosons -- ferbolets
-- are expected as well.

As a final remark, since recombination now requires four particle to
collide, it is not unreasonable to expect rather long lifetimes for
these new objects. Three--body recombination into deep two--body bound
states (if such states exist) could however define the lifetime of
these objects \cite{recomb}.

\begin{figure}[tbh] 

\begin{center} 

\epsfxsize=6.5cm

\centerline{\epsffile{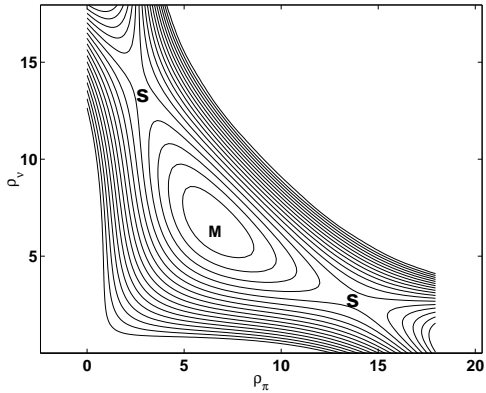}} 

\end{center} 

\caption{ A typical contour plot of the energy density for an
homogeneous Fermi system consisting of two fermion species $\pi$ and
$\nu$, see Eq.(\ref{eq:fermilet}), for the case $\hbar=m=g_3=1$ and
$g_2=-5$.  Only the negative part of the energy density surface is
plotted. The local minimum and the two saddle points are labeled by M
and S respectively. For other sets of parameters there is either only
one saddle point or none at all. }

\label{fig1} 

\end{figure}

I thank G.F. Bertsch, V. Efimov, E.M. Henley and B. Spivak for
discussions, DOE for financial support, N. Takigawa and JSPS for
hospitality and financial support in Sendai, Japan, where
part of this work was performed.


\begin{thebibliography}{99}
\bibitem{scaleout} If one rescales the density by the factor
$|g_2|/g_3$ and the distances by the factor $\sqrt{ \hbar^2
g_3/mg_2^2}$ it is easy to show that the energy of a $N$ particle
boselet is given by the following universal formula: $ E(N)=
\hbar^{3}/ 2m^{3/2}g_3^{1/2} \int d ^3x [
|\bbox{\nabla}\psi|^2 - \rho^2 + \rho^3/3]$ while the number of
particles in a boselet is given by $\hbar^{3}g_3^{1/2}/
m^{3/2}g_2^2 \int d ^3x \rho $.
\end{thebibliography}
\end{document}